\def \AuthorVersion {}

\ifdefined\LaTeXdiff
	\let\VersionWithComments\undefined
	\let\WithReply\undefined
	
\fi

\ifdefined\AuthorVersion
	\newcommand{\VersionLongue}[1]{#1}
\else
	\newcommand{\VersionLongue}[1]{}
\fi

\documentclass[review]{elsarticle}

\usepackage{hyperref} 

\journal{Information Processing Letters}

\usepackage[utf8]{inputenc}

\usepackage{amsmath}
\usepackage{amssymb} 

\usepackage{paralist} 

\usepackage[capitalise,english,nameinlink]{cleveref}

\usepackage{subcaption}

\usepackage{tikz}
\usetikzlibrary{arrows,automata,positioning,shadows,shapes}

\tikzstyle{every node}=[initial text=]
\tikzstyle{location}=[rectangle, rounded corners, minimum size=12pt, draw=black, fill=blue!10, inner sep=2pt]
\tikzstyle{invariant}=[draw=black, dotted, inner sep=1pt] 
\tikzstyle{invisible}=[draw=none]
\tikzstyle{final}=[double]

	\definecolor{coloract}{rgb}{0.50, 0.70, 0.30}
	\definecolor{colorclock}{rgb}{0.4, 0.4, 1}
	\definecolor{colordisc}{rgb}{1, 0, 1}
	\definecolor{colorloc}{rgb}{0.4, 0.4, 0.65}
	\definecolor{colorparam}{rgb}{1, 0.6, 0.0}

\newtheorem{theorem}{Theorem}

\newtheorem{definition}{Definition}

\ifdefined\VersionWithComments
	\usepackage{draftwatermark}
	\SetWatermarkText{draft}
	\SetWatermarkScale{12}
	\SetWatermarkColor[gray]{0.85}
\fi

\definecolor{darkblue}{rgb}{0.0,0.0,0.6}
\definecolor{darkgreen}{rgb}{0, 0.5, 0}
\definecolor{darkpurple}{rgb}{0.7, 0, 0.7}
\definecolor{violetcurie}{RGB}{115,26,67}
\definecolor{forestgreen}{rgb}{0.13,0.54,0.13}
\definecolor{darkblue}{rgb}{0, 0, 0.7}

\ifdefined \VersionWithComments
	\usepackage{marginnote}
	\newcommand{\marginX}{\marginnote{\huge{\quad\quad\textbf{!}\quad\quad}}}
	\newcommand{\ea}[1]{\mbox{}{\color{green!50!black}\marginX{}\textbf{[\'Etienne}: #1]}}
	\newcommand{\lsw}[1]{\mbox{}{\color{orange}\marginX{}\textbf{[Shang-Wei}: #1]}}
	\newcommand{\instructions}[1]{{\color{red}\marginX{}\textbf{[Instructions: ``#1'']}}}
	\newcommand{\reviewer}[2]{\mbox{}{\color{red}\marginX{}\textbf{[Reviewer #1}: ``#2'']}}
	\newcommand{\todo}[1]{\mbox{}{\color{blue}{\marginX{}\textbf{TODO}\ifx#1\\\else:\ \fi #1}}} 
\else
	\newcommand{\instructions}[1]{}
	\newcommand{\ea}[1]{}
	\newcommand{\lsw}[1]{}
	\newcommand{\reviewer}[2]{}
	\newcommand{\todo}[1]{}
\fi

\newcommand{\A}{\ensuremath{\mathsf A}}

\newcommand{\init}{_0}

\newcommand{\Actions}{\Sigma}
\newcommand{\action}{a}

\newcommand{\Clock}{X} 
\newcommand{\clock}{x} 
\newcommand{\clockval}{\ensuremath{\mu}} 
\newcommand{\ClocksZero}{\vec{0}}
\newcommand{\compOp}{\bowtie}

\newcommand{\constantorparameter}{\eta}
\newcommand{\edge}{e}
\newcommand{\Edges}{E}

\newcommand{\fleche}[1]{\stackrel{#1}{\rightarrow}}
\newcommand{\flecheRel}{{\rightarrow}}

\newcommand{\grandn}{{\mathbb N}}
\newcommand{\grandq}{{\mathbb Q}}
\newcommand{\grandqplus}{\grandq_{+}} 
\newcommand{\grandr}{{\mathbb R}}
\newcommand{\grandrplus}{\grandr_{+}} 
\newcommand{\grandz}{{\mathbb Z}}
\newcommand{\guard}{g}
\newcommand{\invariant}{I}
\newcommand{\loc}{l} 
\newcommand{\lochalt}{\ensuremath{\loc_{\mathrm{halt}}}}
\newcommand{\locaccone}{\ensuremath{\loc_{\mathrm{acc}}^1}}
\newcommand{\locacctwo}{\ensuremath{\loc_{\mathrm{acc}}^2}}
\newcommand{\locsink}{\ensuremath{\loc_{\mathrm{sink}}}}
\newcommand{\locstart}{\ensuremath{\loc_{\mathrm{start}}}}
\newcommand{\locinit}{\loc\init}
\newcommand{\Loc}{L} 

\newcommand{\Param}{P} 
\newcommand{\param}{p} 

\newcommand{\pval}{v} 
\newcommand{\sinit}{s\init} 
\newcommand{\state}{s}
\newcommand{\States}{S} 
\newcommand{\UL}{\mathsf{UL}}

\newcommand{\varrun}{\rho} 

\newcommand{\Counteri}[1]{\mathbf{C}_{#1}}
\newcommand{\gadgetTwoCM}{\ensuremath{\A_{2CM}}}
\newcommand{\stateTwoCM}{\mathbf{s}}
\newcommand{\stateTwoCMhalt}{\stateTwoCM_{\mathrm{halt}}}
\newcommand{\StatesTwoCM}{\mathbf{S}}
\newcommand{\TwoCM}{\ensuremath{\mathbf{M}}}

\def\incr#1{\ensuremath{#1++}} 
\def\decr#1{\ensuremath{#1--}} 

\newcommand{\reset}[2]{\ensuremath{[#1]_{#2}}}
\newcommand{\valuate}[2]{\ensuremath{#2(#1)}}

\ifdefined \VersionWithComments
 	\definecolor{colorok}{RGB}{80,80,150}
\else
	\definecolor{colorok}{RGB}{0,0,0}
\fi

\newcommand{\ie}{\textcolor{colorok}{i.\,e.,}} 
\newcommand{\stspace}{\textcolor{colorok}{s.t.}\ } 

\graphicspath{{Figures/}}

\ifdefined\AuthorVersion
\else
\fi

\begin{document}

\begin{frontmatter}

\VersionLongue{%
	\tnotetext[mytitlenote]{This is the author version of the article of the same name published in Information Processing Letters, volume 136, pages 17--20 (2018).
	The final version is available at \href{http://www.dx.doi.org/10.1016/j.ipl.2018.03.013}{10.1016/j.ipl.2018.03.013}.}
}

\title{The language preservation problem is undecidable for parametric event-recording automata%
	\VersionLongue{%
		\tnoteref{mytitlenote}
	}%
}%


\author[EAaddress]{Étienne André\corref{mycorrespondingauthor}}
\ead[url]{https://lipn.univ-paris13.fr/~andre/}

\author[LSWaddress]{Shang-Wei Lin\corref{LSW}}
\cortext[mycorrespondingauthor]{This work is partially supported by the ANR national research program PACS (ANR-14-CE28-0002).} 
\cortext[LSW]{This research is mainly supported by the startup grant M4081588.020.500000 of School of Computer Science and Engineering in Nanyang Technological University.}
\ead{shangweilin@gmail.com}

\address[EAaddress]{Université Paris 13, LIPN, CNRS, UMR 7030, F-93430, Villetaneuse, France}
\address[LSWaddress]{Nanyang Technological University, 50 Nanyang Avenue, Singapore, 639798}

\begin{abstract}
	Parametric timed automata (PTA) extend timed automata with unknown constants (``parameters''), at the price of undecidability of most interesting problems.
	The (untimed) language preservation problem (``given a parameter valuation, can we find at least one other valuation with the same untimed language?'')\ is undecidable for PTAs.
	We prove that this problem remains undecidable for parametric event-recording automata (PERAs), a subclass of PTAs that considerably restrains the way the language can be used; we also show it remains undecidable even for slightly different definitions of the language, \ie{} finite sequences of actions ending in or passing infinitely often through accepting locations, or just all finite untimed words (without accepting locations).
\end{abstract}



\end{frontmatter}


\instructions{\url{https://www.elsevier.com/journals/information-processing-letters/0020-0190/guide-for-authors}}

\section{Introduction}\label{section:introduction}

Timed automata (TAs)~\cite{AD94} are a useful formalism to model and formally verify systems involving timing hard constraints and concurrency.
TAs benefit from numerous decidability results, including the emptiness of the accepted language.
However, the universality and the language inclusion are undecidable for timed automata~\cite{AD94}.
Therefore, subclasses have been proposed.
The language inclusion becomes decidable for event-recording automata (ERAs)~\cite{AFH99}.

Parametric timed automata (PTAs)~\cite{AHV93} extend TAs with timing parameters: this very expressive formalism can model systems where timing constants are uncertain or unknown, at the cost of most interesting problems to be undecidable~\cite{Andre17}.
The mere emptiness of the valuation set for which a given location is reachable (``reachability-emptiness'') is undecidable~\cite{AHV93}.

Restricting the syntax of a formalism may bring decidability: the language inclusion undecidable for TAs~\cite{AD94} becomes decidable for ERAs~\cite{AFH99}.
In contrast, the reachability emptiness problem for PTAs remains undecidable for a subclass of PTAs with only open inequalities~\cite{Doyen07}.

In~\cite{ALin17}, we proposed parametric event-recording automata (PERAs), and showed that the reachability-emptiness problem remains undecidable for PERAs.
Although it seems that our proof idea can be extended to most problems where the language (\ie{} the transition labels) does not play a role (which would include unavoidability-emptiness~\cite{JLR15}), it remains open whether language problems undecidable for PTAs become or not decidable for PERAs.
In~\cite{AM15}, we showed that the following language preservation problem is undecidable for PTAs: ``given a PTA and a reference parameter valuation, does there exist another valuation with the same untimed language?''.
This problem has connections with the \emph{robustness} of timed systems, as it asks whether other valuations of the timing constants may lead to the same discrete behavior.
The set of valuations with the same untimed language can also be used to perform optimization of some constants without impacting the system's (untimed) behavior.
\label{newtext:motivation}

We show here that the language preservation problem is undecidable for PERAs, and remains undecidable for different definitions of the language.\ea{and what about L/U-PERAs?}
This quite surprising result comes in contrast with the difference of decidability between TAs and ERAs in the non-parametric setting.

\section{Preliminaries}\label{section:preliminaries}

\VersionLongue{
\subsection{Clocks, parameters and constraints}
}

\VersionLongue{
Let $\grandn$, $\grandz$, $\grandqplus$ and $\grandrplus$ denote the sets of non-negative integers, integers, non-negative rational and non-negative real numbers respectively.
}

Throughout this paper, we assume a set~$\Clock$ of \emph{clocks}, \ie{} real-valued variables that evolve at the same rate.
A clock valuation is a function
$\clockval : \Clock \rightarrow \grandrplus$.
We write $\ClocksZero$ for the clock valuation that assigns $0$ to all clocks.
Given $d \in \grandrplus$, $\clockval + d$ denotes the valuation such that $(\clockval + d)(\clock) = \clockval(\clock) + d$, for all $\clock \in \Clock$.
Given $\clock \in \Clock$, we define the \emph{reset} of a valuation~$\clockval$, denoted by $\reset{\clockval}{\clock}$, as follows: $\reset{\clockval}{\clock}(\clock') = 0$ if $\clock' = \clock$, and $\reset{\clockval}{\clock}(\clock')=\clockval(\clock')$ otherwise.

We assume a set~$\Param$ of \emph{parameters}, \ie{} unknown rational-valued constants.
A parameter {\em valuation} $\pval$  is a function
$\pval : \Param \rightarrow \grandqplus$.

A \emph{guard}~$\guard$ is a constraint over $\Clock \cup \Param$ defined by a conjunction of inequalities of the form
	$\clock \compOp \constantorparameter$, where $\constantorparameter$ is either a parameter or a constant in~$\grandz$, and ${\compOp} \in \{<, \leq, \geq, >\}$.
Given~$\pval$, $\valuate{\guard}{\pval}$ denotes~$\guard$ where all occurrences of each parameter~$\param_i \in \Param$ have been replaced by~$\pval(\param_i)$.

\VersionLongue{
\subsection{Parametric Event-Recording Automata}\label{ss:PTAs}

Parametric event-recording automata (PTAs) extend event-recording automata with parameters within guards and invariants in place of integer constants~\cite{ALin17}; they can also be seen as a syntactic subclass of parametric timed automata~\cite{AHV93}.
}

\begin{definition}[PERA]\label{def:PTA}
	A \emph{parametric event-recording automaton} (\VersionLongue{hereafter }PERA)
	$\A$ is a tuple \mbox{$(\Actions, \Loc, \locinit, \Param, \invariant, \Edges)$}, where: 
	\begin{inparaenum}[\itshape i\upshape)]%
		\item $\Actions$ is a finite set of actions,
		\item $\Loc$ is a finite set of locations,
		\item $\locinit \in \Loc$ is the initial location,
		\item $\Param$ is a finite set of parameters,
		\item $\invariant$ is the invariant, assigning to every $\loc\in \Loc$ a guard $\invariant(\loc)$,
		\item $\Edges$ is a finite set of edges  $\edge = (\loc,\guard,\action,\loc')$
		where
        $\loc,\loc'\in \Loc$ are the source and target locations, $\action \in \Actions$, 
        and $\guard$ is a guard.
	\end{inparaenum}
\end{definition}

A PERA has a one-to-one mapping between clocks and actions.
Given\VersionLongue{ a set of actions}~$\Actions$, let $\Clock_\Actions$ denote the set of associated clocks.
Similarly, we denote by $\clock_\action$ the clock associated with~$\action$.
On any transition labeled with~$\action$, $\clock_\action$ is implicitly reset.

Given a PERA~$\A$ and a parameter valuation $\pval$, we denote by $\valuate{\A}{\pval}$ the non-parametric event-recording automaton where all occurrences of a parameter~$\param_i$ have been replaced by~$\pval(\param_i)$, for each $\param_i \in \Param$.


\begin{definition}[Concrete semantics\VersionLongue{ of an ERA}]
	Given a PERA $\A = (\Actions, \Loc, \locinit, \Param, \invariant, \Edges)$, 
	and a parameter valuation~\(\pval\),
	the concrete semantics of $\valuate{\A}{\pval}$ is given by the timed transition system $(\States, \sinit, \flecheRel)$, with
	$\States = \{ (\loc, \clockval) \in \Loc \times \grandrplus^{|\Clock_\Actions|} \mid  
					\clockval \models \valuate{\invariant(\loc)}{\pval}
					\}$%
            , $\sinit = (\locinit, \ClocksZero) $,
            and
	$\flecheRel$ consists of the discrete and \VersionLongue{(continuous) }delay transition relations:
				\begin{itemize}
			\item discrete\VersionLongue{ transitions}: $(\loc,\clockval) \fleche{\edge} (\loc',\clockval')$, 
				if $(\loc, \clockval) , (\loc',\clockval') \in \States$, there exists $\edge = (\loc,\guard,\action,\loc') \in \Edges$, $\clockval'= \reset{\clockval}{\clock_\action}$, and
					$\clockval \models \valuate{\guard}{\pval}$.
			\item delay\VersionLongue{ transitions}: $(\loc,\clockval) \fleche{d} (\loc, \clockval+d)$, with $d \in \grandrplus$, 
				if \mbox{$\forall d' \in [0, d], (\loc, \clockval+d') \in \States$.}
		\end{itemize}
\end{definition}


A \emph{run} is a sequence $\varrun=\state_0 \gamma_0 \state_1 \gamma_1 \cdots \state_n \gamma_n \cdots$ such that $\forall i, \state_i \fleche{\gamma_i} \state_{i+1}$.
We consider as usual that runs strictly alternate delays $d_i$ and discrete transitions $\edge_i$ and we thus write runs in the form $\varrun = \state_0 \fleche{(d_0,\edge_0)} \state_1 \fleche{(d_1,\edge_1)}\cdots$.
The corresponding \emph{timed word} is $(\action_0, t_0), (\action_1, t_1), \cdots$ where $\action_i$ is the action of~$\edge_i$ and $t_i = \sum_{j=0}^i d_i$.
The corresponding \emph{untimed word} is $\action_0 \action_1 \cdots$.
A maximal run is a run that is either infinite, or that cannot be extended by a discrete transition.
\VersionLongue{

}As in~\cite{AM15}, we define the \emph{language} of $\valuate{\A}{\pval}$, denoted by~$\UL(\valuate{\A}{\pval})$ as the set of all untimed words associated with a maximal run of~$\A$.


%

\section{Encoding a 2-counter machine into a PERA}\label{section:enconding}

\lsw{I suppose locations are used in PERA, while states are use in 2-counter machines. However, I found they are used interchangeably. }\ea{Yes you are right. Oops! I double checked and corrected some occurrences.}

We propose in this section an encoding of a 2-counter machine (2CM) into a PERA.
This encoding is adapted from the PTA encoding of~\cite{AM15} to our setting of PERAs, and is therefore not a main contribution of this work.

Fix a deterministic 2CM~$\TwoCM$. 
Recall that such a machine has two non-negative counters $\Counteri{1}$ and $\Counteri{2}$ (the value of which is initially~0), and a finite number of states and of transitions of the form:
\begin{itemize}
	\item when in state $\stateTwoCM_i$, increment $\Counteri{k}$ and go to $\stateTwoCM_j$;
	\item when in state $\stateTwoCM_i$, if $\Counteri{k} = 0$ then go to $\stateTwoCM_j$ else decrement $\Counteri{k}$ and go to $\stateTwoCM_{j'}$.
\end{itemize}

The machine starts in state $\stateTwoCM_0$ and halts when it reaches a particular state $\stateTwoCMhalt{}$.
The halting problem for 2-counter machines is undecidable \cite{Minsky67}.

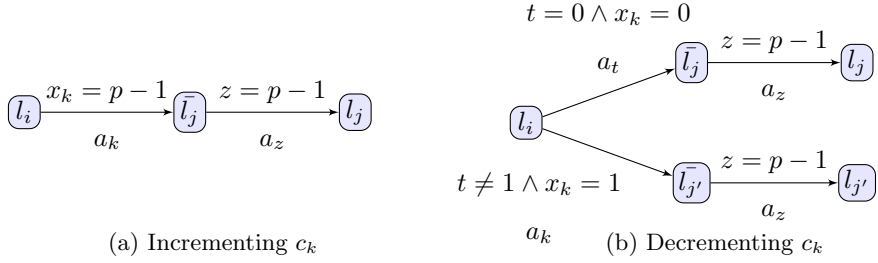
\begin{figure} [t]
\centering

\begin{subfigure}[b]{0.45\textwidth}
	\begin{tikzpicture}[-latex',xscale=1.1]

		\node[location] (s) at(0,0) {$\loc_i$};
		\node[location] (s'barre) at(2,0) {$\bar{\loc_j}$};
		\node[location] (s') at(4,0) {$\loc_j$};

		\path
			(s) edge node {\begin{tabular}{c}
					$x_k=\param-1$\\
					$\action_k$\\
				\end{tabular}} (s'barre)
			(s'barre) edge node{\begin{tabular}{c}
					$z = \param-1$\\
					$\action_z$ \\
				\end{tabular}} (s')
			;
					\node[location, draw=none,fill=none,use as bounding box] at (4,-1) {};
	\end{tikzpicture}
	\caption{Incrementing $c_k$}
	\label{figure:reduction:increment}
\end{subfigure}%
\hfill
\begin{subfigure}[b]{0.45\textwidth}
	\begin{tikzpicture}[-latex',xscale=1.1]

		\node[location] (s) at(0,0) {$\loc_i$};
		\node[location] (s0barre) at(2,.8) {$\bar{\loc_j}$};
		\node[location] (s0) at(4,.8) {$\loc_j$};
		\node[location] (s1barre) at(2,-.8) {$\bar{\loc_{j'}}$};
		\node[location] (s1) at(4,-.8) {$\loc_{j'}$};
					\path[use as bounding box] (0,0);
		\path
			(s) edge node[above] {\begin{tabular}{c}
					$t = 0 \land x_k = 0$\\
					$\action_t$
				\end{tabular}} (s0barre)
			(s) edge node[below left=-7pt,pos=.7] {\begin{tabular}{c}
					$t\not=1 \land x_k = 1$\\
					$\action_k$ 
				\end{tabular}} (s1barre)
			(s0barre) edge node {\begin{tabular}{c}
					$z = \param-1$\\
					$\action_z$ 
				\end{tabular}} (s0)
			(s1barre) edge node {\begin{tabular}{c}
					$z = \param-1$\\
					$\action_z$ 
				\end{tabular}} (s1)
			;
	\end{tikzpicture}
	\caption{Decrementing $c_k$}
	\label{figure:reduction:decrement}
\end{subfigure}%

\caption{Encoding a 2-counter machine into a PERA}
\label{figure:reduction}
\end{figure}

Our encoding requires a single parameter~$\param$ and four actions $\action_t$, $\action_1$, $\action_2$, and~$\action_z$ (associated with clocks $t$, $x_1$, $x_2$ and~$z$ respectively).
Clock~$t$ serves as a tick (it is reset exactly every $\param$ time units).
We encode a configuration of the 2CM as follows: whenever~$t=0$, then $x_1=c_1$ and~$x_2=c_2$, where $c_1,c_2$ are the current values of $\Counteri{1},\Counteri{2}$.
Finally, $z$ is used to count the number of steps of the 2CM: we use~$\param$ to bound the
length of the computation of the 2CM.
%
The PERA~$\A$ associated with~$\TwoCM$ is defined as follows:
\begin{itemize}
	\item its set of locations has two copies of the set~$\StatesTwoCM$ of states of~$\TwoCM$: for each~$\stateTwoCM_i \in \StatesTwoCM$, there is a \emph{main location}~$\loc_i$, and an \emph{intermediary location} named~$\bar{\loc_i}$;
	
	\item Each location of~$\A$ carries three self-loops, associated with each of
	the three clocks~$t$, $x_1$, and~$x_2$, and resetting that clock when it reaches value~$\param$, \ie{} associated with actions~$\action_t$, $\action_1$, and~$\action_2$ respectively.\ea{example of self-loop?}
	This requires a global invariant enforcing that all four clocks~$t$, $x_1$, $x_2$, and~$z$ remain below~$\param$.

	Then each transition~$(\stateTwoCM_i, \incr{c_k}, \stateTwoCM_j)$ incrementing counter~$\Counteri{k}$
	in~$\TwoCM$ gives rise to a transition from location~$\loc_i$ to~$\bar{\loc_j}$,
	with guard~$x_k=\param-1$, and labeled with~$\action_k$ (therefore resetting $x_k$)
	(see \cref{figure:reduction:increment}).
	Each transition of the
	form~$(\stateTwoCM_i, \decr{c_k}, \stateTwoCM_j, \stateTwoCM_{j'})$ is handled similarly, but gives rise to two
	transitions: one transition from~$\loc_i$ to~$\bar{\loc_j}$ with guard~$t=0 \land x_k=0$ (the actual 0-test for~$\Counteri{k}$), and one transition from~$\loc_i$ to~$\bar{\loc_{j'}}$ with guard $x_k=1$ and labeled with~$\action_k$ (therefore resetting~$x_k$).
	Then, from each location~$\bar{\loc}$ of~$\A$, there is
	a transition to the corresponding location~$\loc$ with guard~$z=\param-1$ and
	resetting~$z$ due to action~$\action_z$ (see \cref{figure:reduction:decrement}).
\end{itemize}

Clock~$z$ counts the number of steps (when considering the value of this clock when~$t=0$, it~encodes a counter that~is incremented at every transition of~$\TwoCM$).
\VersionLongue{Notice that clock~$z$ counts, but for the moment, it~does not impose any constraint on the length of the simulation.
}%
%
Let us now add condition $0<t<\param$ 
to the guards
$z=\param-1$ of the transitions leaving the intermediary locations.
This way, when $z$ (seen as a counter) has value~$\param-1$ (when $t = 0$ or $\param$), no~transition
is available from any location~$\bar s$, 
so that the encoding stops after mimicking $\param-1$ steps of the execution of~$\TwoCM$.
As a consequence, our encoding~$\valuate{\A}{\pval}$ only encodes properly the~$\pval(\param)$ first steps of~$\TwoCM$, and then blocks (therefore steps beyond $\pval(\param)$ steps are not encoded at all).

\section{Undecidability of the language preservation}\label{section:undecidability}

\subsection{Main result}

\VersionLongue{%
We now show our main result below.
}

\begin{theorem}\label{theorem:undecidability}
	The language preservation problem for PERAs is undecidable.
\end{theorem}

The proof of undecidability of~\cite{AM15} for PTAs strongly relies on the fact that all transitions were labeled with the same action~$\action$.
This reasoning cannot be kept here, as the transitions of our modified encoding in \cref{section:enconding} are labeled with different actions so as to reset different clocks.
Therefore, it is not possible to know in advance the language of the accepting run of the 2CM (if any).

For example, assume a run of the 2CM made of the following two instructions:
when in state $\stateTwoCM_0$, increment $\Counteri{1}$ and go to $\stateTwoCM_3$;
when in state $\stateTwoCM_3$, if $\Counteri{2} = 0$ then go to $\stateTwoCM_5$ else decrement $\Counteri{2}$ and go to~$\stateTwoCM_0$.
The sequence of states and actions in our PERA will be
$(\loc_0,\clockval_0)
\fleche{\action_1}
(\bar{\loc_3},\clockval_1)
\fleche{\action_z}
(\loc_3,\clockval_2)
\fleche{\action_t}
(\bar{\loc_5},\clockval_3)
\fleche{\action_z}
(\loc_5,\clockval_4)
$, for some $\clockval_0,\cdots,\clockval_4$.
(In this sequence, we write $\action_i$ instead of $(d_i, \edge_i)$ to make clear the action~$\action_i$ used in edge~$\edge_i$).
Therefore, the untimed language corresponding to these two instructions is $\action_1 \action_z \action_t \action_z$.

In order to prove \cref{theorem:undecidability}, we reduce from the halting problem of a 2CM, using the encoding of \cref{section:enconding}.
We will allow all possible infinite \VersionLongue{(untimed) }words for the reference valuation, and will rely on the difference between finite and infinite words to perform a distinction between the halting or the non-halting case.

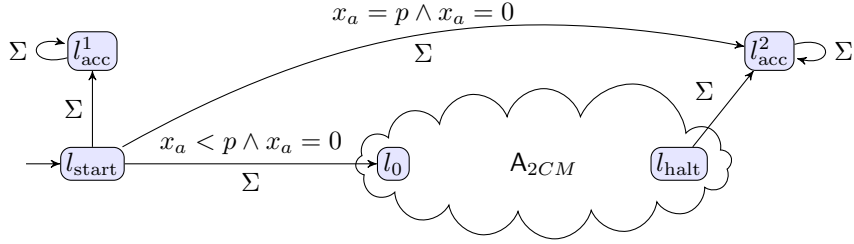
\begin{figure}[tb]
	{\centering
	\begin{tikzpicture}[->, >=stealth', auto, node distance=2cm, thin]

		\node[location, initial] (lstart) at (-6, 0) {$\locstart$};
		\node[location] (l0) at (-2, 0) {$\loc_0$};
		\node[location] (lh) at (+1.8, 0) {$\lochalt$};
		\node[location] (lacc1) at (-6, 1.5) {$\locaccone$};
		\node[location] (lacc2) at (+3, 1.5) {$\locacctwo$};
		
		\node[cloud, cloud puffs=15.7, cloud ignores aspect, minimum width=5cm, minimum height=2cm, align=center, draw] (cloud) at (0cm, 0cm) {$\gadgetTwoCM$};
		

		\path
			(lstart) edge[] node[above] {$\clock_a < \param \land \clock_a  = 0$} node[below] {$\Actions$} (l0)
			(lstart) edge[] node {$\Actions$} (lacc1)
			(lstart) edge[bend angle=20,bend left] node[above] {$\clock_a = \param \land \clock_a  = 0$} node[below]{$\Actions$} (lacc2)
			(lh) edge[] node {$\Actions$} (lacc2)
			(lacc1) edge[loop left] node[] {$\Actions$} (lacc1)
			(lacc2) edge[loop right] node[] {$\Actions$} (lacc2)
			;

	\end{tikzpicture}
	
	}
	\caption{Undecidability of the language preservation problem for PERAs}
	\label{fig:proof:undecidability}
\end{figure}

Our proof relies on the PERA~$\A$ given in \cref{fig:proof:undecidability}, that contains the encoding of a 2CM~$\TwoCM$, denoted by~\gadgetTwoCM{}.
A transition labeled with~$\Actions$ denotes 4 transitions labeled with $\action_t, \action_1, \action_2, \action_z$ respectively.
Let~$\pval_0$ be the valuation \stspace{}$\pval_0(\param) = 0$.
Given~$\A$ and~$\pval_0$, we will show that there exists $\pval \neq \pval_0$ \stspace{}$\UL(\valuate{\A}{\pval_0}) = \UL(\valuate{\A}{\pval})$ iff $\TwoCM$ halts.
First, let us study $\valuate{\A}{\pval_0}$: this TA can take the transitions to either $\locaccone$ (which is ungarded) or~$\locacctwo$ (which requires $\param = 0$), but not that to~$\loc_0$ as the guard requires $\param > 0$.
Therefore, $\UL(\valuate{\A}{\pval_0}) = ( \action_t | \action_1 | \action_2 | \action_z )^\omega$, \ie{} the language made of exactly all infinite words, hereafter~$\Actions^\omega$.

Now, assume the machine halts after $n$ steps.
There exists a parameter valuation $\pval$ (typically \stspace{}$\pval(\param) > n$) \stspace{}the machine is correctly simulated.
The (unique) run going through the gadget \gadgetTwoCM{} is non-blocking and reaches location~\lochalt{}, from which it goes to~$\locacctwo$ and can perform any action an infinite number of times.
The corresponding possible runs are therefore included into~$\Actions^\omega$.
Since this valuation can also take the transition from \locstart{} to \locaccone{}, then the language is $\Actions^\omega$.
Hence there exists $\pval \neq \pval_0$ \stspace{}$\UL(\valuate{\A}{\pval_0}) = \UL(\valuate{\A}{\pval})$.

Assume the machine does not halt, and consider any valuation $\pval \neq \pval_0$.
As in the previous case, the transition from \locstart{} to \locaccone{} can be taken, giving (at least) $\Actions^\omega$.
In addition, for this valuation, the transition to~$\loc_0$ can be taken (since $\pval(\param) > 0$), and the 2CM starts to be simulated.
However, from our encoding in \cref{section:enconding}, this run will stop after~$\pval(\param)$ steps, and will block (without reaching~\lochalt{} as the 2CM does not halt).
This blocking run is a finite blocking run, therefore is a maximal run, and is therefore part of the language.
Hence the language contains all infinite runs ($\Actions^\omega$) plus one finite blocking run---which was not part of~$\valuate{\A}{\pval_0}$.
Therefore, for all $\pval \neq \pval_0$, $\UL(\valuate{\A}{\pval_0}) \subsetneq \UL(\valuate{\A}{\pval})$.

This gives that there exists $\pval \neq \pval_0$ \stspace{}$\UL(\valuate{\A}{\pval_0}) = \UL(\valuate{\A}{\pval})$ iff $\TwoCM$ halts.
\hfill$\blacksquare$

\subsection{Varying the definition of language}

We can wonder whether the undecidability comes from our definition of the language (consistent with~\cite{AM15}).
We briefly discuss other cases to show that it does not.
To prove undecidability of the three cases below, we need to perform a common modification to the 2CM encoding.
From any intermediary\lsw{intermediary?}\ea{I'm not sure} location $\bar{\loc}$ of \gadgetTwoCM{}, we add a transition guarded with $t = \param \land z = \param - 1$ leading to a new location~$\locsink{}$ with actions~$\Actions$.
This transition can only be taken after exactly~$\pval(\param)$ steps of the 2CM, \ie{} instead of blocking after $\pval(\param)$ steps, the run goes to~\locsink{}.

\subsubsection{Büchi condition or reachability condition}

Let us redefine the untimed language as the set of all words associated with a run passing infinitely often through an accepting location.
In that case, our scheme in \cref{fig:proof:undecidability} can be kept with only mild modifications:
let \locaccone{}, \locacctwo{} and \locsink{} be the accepting locations.
Let us add a self-loop on \locsink{} with a new action, say~$\action_3$.
For~$\pval_0$, the untimed language is $\Actions^\omega$ (the notation $\Actions$ remains unchanged, \ie{} does not contain~$\action_3$).
If the 2CM halts, some valuations will reach \lochalt{}, and the untimed language is $\Actions^\omega$.
If the 2CM does not halt, $\Actions^\omega$ is part of the language but, for all $\pval \neq \pval_0$, the run will block, and end in \locsink{} where it will perform an infinite word with suffix $(\action_3)^\omega$, which differs from the discrete behavior of~$\valuate{\A}{\pval_0}$.\label{newtext:typo}

%
The case of the language defined as the set of finite words ending in an accepting location is similar.

\subsubsection{Safety untimed language}\label{sss:safety}

Finally, let us redefine the untimed language as the set of untimed words associated with any finite run (no accepting locations are considered).
In that case, we relabel the transitions to~$\locsink$ with the fresh action~$\action_3$.
For~$\pval_0$, the language becomes $\Actions^*$.
If the 2CM halts, the language is again~$\Actions^*$ for some valuations.
However, if the 2CM does not halt, for $\pval \neq \pval_0$ the (unique) run will block, and end in $\locsink$ with $\action_3$ as suffix---yielding a word not part of $\UL(\valuate{\A}{\pval_0})$.

\section{Conclusion}\label{section:conclusion}

We proved that the language preservation remains undecidable for a subclass of PTAs, namely parametric event-recording automata\lsw{it looks like the subclass of PERA, but actually, we mean PERA, the subclass of PTA, right?}.
We believe that the L/U-automata restrictions considered in the additional undecidability results of~\cite{AM15} could apply to our setting, and undecidability would still hold for ``L/U-PERAs''.
A more challenging future work is to study the trace preservation problem of~\cite{AM15} that considers not only the actions but also the locations.

\ifdefined\AuthorVersion
	\newcommand{\LNCS}{Lecture Notes in Computer Science}
	\newcommand{\STTT}{International Journal on Software Tools for Technology}
	\newcommand{\TCS}{Theoretical Computer Science}
	\newcommand{\TSE}{IEEE Transactions on Software Engineering}
\else
	\section*{References}
	\newcommand{\LNCS}{LNCS}
	\newcommand{\STTT}{STTT}
	\newcommand{\TCS}{TCS}
	\newcommand{\TSE}{IEEE TSE}
\fi
\bibliography{lgPERA}

\ifdefined\WithReply
	\input{letter1.tex}
\fi

\end{document}